\pdfoutput=1
\documentclass[twocolumn,showpacs,amssymb,floatfix,aps]{revtex4}

\usepackage{graphicx}
\usepackage{dcolumn}
\usepackage{bm}
\usepackage{epsfig}

\begin{document}
CERN-PH-TH/2010-202
\preprint{CERN-PH-TH/2010-202}
\title{
Comment on ``The third Zemach moment of the proton", by
Cl$\ddot{\rm o}$et and Miller.}
\author{A. De R\'ujula${}^{a,b,c}$}
\affiliation{  \vspace{3mm}
${}^a$ Instituto de F\'isica Te\'orica, Univ. Aut\'onoma de Madrid, Madrid, and 
CIEMAT, Madrid, Spain,\\
${}^b$ Physics Dept., Boston University, Boston, MA 02215,\\
${}^c$Physics Department, CERN, CH 1211 Geneva 23, Switzerland}

\date{\today}

\begin{abstract}
Cl$\ddot{\rm o}$et and Miller, in arXiv:1008.4345,  state that 
``existing data rule out a value of the third Zemach moment
large enough to explain the current puzzle with the proton
charge radius determined from the Lamb shift in muonic
Hydrogen. This is in contrast with
the recent claim of De R\'ujula". 
To be more precise: it is not. It is, however, contrary to what 
they claim that I claim. Cl$\ddot{\rm o}$et and Miller have
simply misinterpreted my claims \cite{Dic}.
\end{abstract}


\maketitle

\section{My main claim}

Let ``model-independent" mean ``independent of a particular parametrization
of the charge distribution of the proton, $\rho_p(r)$''.
The most precise model-independent
 measurements of the corresponding mean square radius,
 $\langle r_p^2\rangle$, are
mainly based on the theory \cite{HTheory} and observations  \cite{Hexps} of  
Hydrogen. The 
result, compiled in CODATA \cite{CODATA}, is
\begin{equation}
{\langle r_p^2\rangle}\rm (CODATA)=(0.8768 \pm 0.0069\; \rm fm)^2
\label{rCODATA}
\end{equation}

Consider next the
 $\rm 2P_{3/2}^{F=2}\to 2S_{1/2}^{F=1}$ Lamb shift in a $\mu p$ atom.
In meV units for energy and fermi units for the radii, the predicted value 
\cite{LymanTH}
is of the form
\begin{eqnarray}
L^{\rm th}\left[\langle r_p^2\rangle,\langle r_p^3 \rangle_{(2)}\right]&=&\nonumber\\
209.9779(49)&-&5.2262\, \langle r_p^2\rangle +0.00913 \,\langle r_p^3 \rangle_{(2)},
\label{Lth}
\end{eqnarray}
which is also model-independent.
The first two coefficients are best  estimates of many contributions \cite{Pohl}
 while the third stems from the $n=2$ value of an addend
  \cite{Friar}
 \begin{equation}
 \Delta E_3(n)= {\alpha^5\over 3\,n^3}\,m_r^4 \,
 \delta_{l0}\,\langle r_p^3 \rangle_{(2)},
 \label{DeltaE3}
 \end{equation}
 proportional to the third Zemach moment
\begin{equation}
\langle r_p^3 \rangle_{(2)}\equiv\int d^3 r_1 d^3 r_2\,\rho(r_1)\rho(r_2)
\vert {\bold r}_1-{\bold r}_2\vert^3
\end{equation}
  
The quoted Lamb shift has been measured \cite{Pohl} to be
\begin{equation}
L_{\rm exp} = 206.2949 \pm 0.0032\;\rm meV.
\label{Lexp}
\end{equation}

Assume that the theory and experiments quoted so
far are correct.
Use Eqs.~(\ref{Lth},\ref{Lexp}) to write $L^{\rm th}=L_{\rm exp}$.
Introduce into the expression for $L^{\rm th}$ the observed value of 
$\langle r_p^2\rangle$ given in Eq.~(\ref{rCODATA}). Solve for
$\langle r_p^3 \rangle_{(2)}$. The result is:
\begin{equation} 
\left[ \langle r_p^3 \rangle_{(2)} \right]^{1/3}=3.32\pm0.22\;\rm fm
\label{finalresult}
\end{equation}
with the error dominated by the CODATA uncertainty.

My main claim is to have done the algebra leading to Eq.~(\ref{finalresult}),
which is model-independent.

\section{My alleged claim}

Cl$\ddot{\rm o}$et and Miller state ``In the simple monopole model of De R\'ujula,
which claims to account for the proton radius puzzle...".

I have not claimed that a monopole model (which, incidentally, is not ``mine")
accounts for the proton radius puzzle. On the contrary, I explicitly constructed
a ``toy model" with two added monolope form factors to illustrate how the
non-vanishing contributions of no fewer than two monopoles were needed to
reconcile the data.

The comment that may have led Cl$\ddot{\rm o}$et and Miller to misinterpret
me is the last paragraph in arXiv:1008.3861:

``The result Eq.~(\ref{finalresult}) is $\rho_p(r)$-independent and to
 be treated with due respect. Right after offering
 excuses, I shall break this rule.
The third Zemach moment is very sensitive to the long-distance part of
$\rho(r)$...
A correct question to ask is the scale, $m$, to which this tail
corresponds. Not to be confused with the toy model fit to both
 $\langle r_p^2\rangle$ and $\langle r_p^3 \rangle_{(2)}$,
the simplest naive --but consistent-- answer
is provided by Eq.~(\ref{finalresult}) in
the single-pole approximation 
i.e.~$m\!\simeq\! 261$ MeV, tantalizingly close to the threshold of the proton
form factor's cut at $2\,m_{\pi^\pm}\!\simeq\! 278$ MeV".

All I did in the previous paragraph is to use a monopole to extract
a rough estimate of the slope of $\rho_p(r)$ at large $r$, and to
express it as an inverse mass. I did not contend that a single
monopole representation of $G_E(-{\bold q}^2)$
``accounts for the proton radius puzzle" at all relevant
values of ${\bold q}^2$, nor that its Fourier transform
accurately describes $\rho_p(r)$ at all relevant $r$.

The penultimate paragraph above, from v2 of arXiv:1008.3861
(posted one day after v1) may have been more obscure in v1.
No doubt this is what mislead
Cl$\ddot{\rm o}$et and Miller, regarding what my claims really were.

\section{Cl$\bf{\ddot{\rm O}}$et and Miller's claims}

I use the word ``claims" here for literary consistency. 
But I mean ``statements".

Cl$\ddot{\rm o}$et and Miller study the proton-radius
puzzle by use of three models --the dipole and the ones in
\cite{Kelly,Alberico}-- that describe  
$G_E({q}^2)$ as measured in $ep$
scattering experiments. They find that the model
for which $G_E(q^2)$ falls fastest at large
$q^2$ gives the largest value of 
$ \langle r_p^3 \rangle^2_{(2)} / \langle r_p^2\rangle^3$,
but is insufficient to resolve the ``puzzle".

In the models in \cite{Kelly, Alberico}
the numerical parameters can be combined with their single
mass scale ($m_p$) to be reinterpreted as different mass scales. 
This is in agreement with a claim that I have indeed made:
``Any simple one-parameter description of the proton's non-relativistic
Sacks form factor, $G_E(-{\bold q}^2)$, in terms of only one mass 
parameter is inaccurate: the proton is not so simple". 

The crucial problem was adroitly emphasized by Sick \cite{Sick}. 
It is very difficult to extract reliable information on $\rho(r)$ 
(such as its moments) from its Fourier
transform, $G_E({\bold q})^2$.
The radius of convergence
of the expansion in $r$ from which one {\it directly} extracts $\langle r_p^2\rangle$ 
from the $ep$ data is so small, that one must use numerical simulations 
and a continued-fraction expansion 
to skirt the uncertainties associated with data normalization at small ${\bold q}^2$
(and their systematic errors)
 and to obtain a stable, numerically-meaningful
result not contaminated, for instance, by the term in $\langle r_p^4\rangle$.
Clearly, if extracting $\langle r_p^2\rangle$ is delicate, the more so it is
to infer $\langle r_p^3 \rangle_{(2)}$. 

The authors of \cite{Alberico} are
aware of the above-mentioned practicalities. But they do not need to face
them \`a la Sick, since they analyze data at much larger
$q^2$ than the range with which we are concerned.
Cl$\ddot{\rm o}$et and Miller simply extrapolate the high-$q^2$
models in \cite{Kelly, Alberico}
to the $|{\bold q}|$ values of ${\cal{O}}(\alpha \, m_\mu)$ of interest here.
In this way, not surprisingly, they
 obtain a model- and extrapolation-dependent 
value of $\langle r_p^3 \rangle^2_{(2)}$
in disagreement with Eq.~(\ref{finalresult}).

\subsection*{Acknowledgments}
I am indebted to York Schroeder and  to Ian
Cl$\ddot{\rm o}$et and Gerald Miller for having independently found a mistake 
in my explicit $\langle r^3\rangle_{(2)}$ numerical expressions, which I shall
correct. This error does not alter my claims.


\begin{thebibliography}{999}
\bibitem{Dic}
``To claim: to state or assert that something is the case, 
typically without providing evidence or proof."
\end{thebibliography}
\end{document}